\documentclass[manuscript,nonacm]{acmart}

\usepackage{algorithm2e}
\usepackage{subcaption}

\AtBeginDocument{%
  }

\setcopyright{cc}
\setcctype{by-sa}
\copyrightyear{2025}

\theoremstyle{definition}

\theoremstyle{definition}
\newtheorem{criterion}{C\ignorespaces}
\makeatletter
\newcommand{\criterionautorefname}{C\@gobble}
\makeatother

\begin{document}

\title{VEIL: Reading Control Flow Graphs Like Code}

\author{Philipp Schaad}
\orcid{0000-0002-8429-7803}
\email{philipp.schaad@inf.ethz.ch}
\affiliation{%
  \institution{ETH Zurich}
  \city{Zurich}
  \country{Switzerland}
}

\author{Tal Ben-Nun}
\orcid{0000-0002-3657-6568}
\email{talbn@llnl.gov}
\affiliation{
    \institution{Lawrence Livermore National Laboratory (LLNL)}
    \city{Livermore}
    \country{USA}
}

\author{Torsten Hoefler}
\orcid{0000-0002-1333-9797}
\email{htor@inf.ethz.ch}
\affiliation{
    \institution{ETH Zurich}
    \city{Zurich}
    \country{Switzerland}
}

\renewcommand{\shortauthors}{Schaad et al.}

\begin{abstract}
Control flow graphs (CFGs) are essential tools for understanding program behavior, yet the size of real-world CFGs makes them difficult to interpret.
With thousands of nodes and edges, sophisticated graph drawing algorithms are required to present them on screens in ways that make them readable and understandable.
However, being designed for general graphs, these algorithms frequently break the natural flow of execution, placing later instructions before earlier ones and obscuring critical program structures.
In this paper, we introduce a set of criteria specifically tailored for CFG visualization, focusing on preserving execution order and making complex structures easier to follow.
Building on these criteria, we present VEIL, a new layout algorithm that uses dominator analysis to produce clearer, more intuitive CFG layouts.
Through a study of CFGs from real-world applications, we show how our method improves readability and provides improved layout performance compared to state of the art graph drawing techniques.
\end{abstract}



\begin{teaserfigure}
    \centering
    \includegraphics[width=\linewidth]{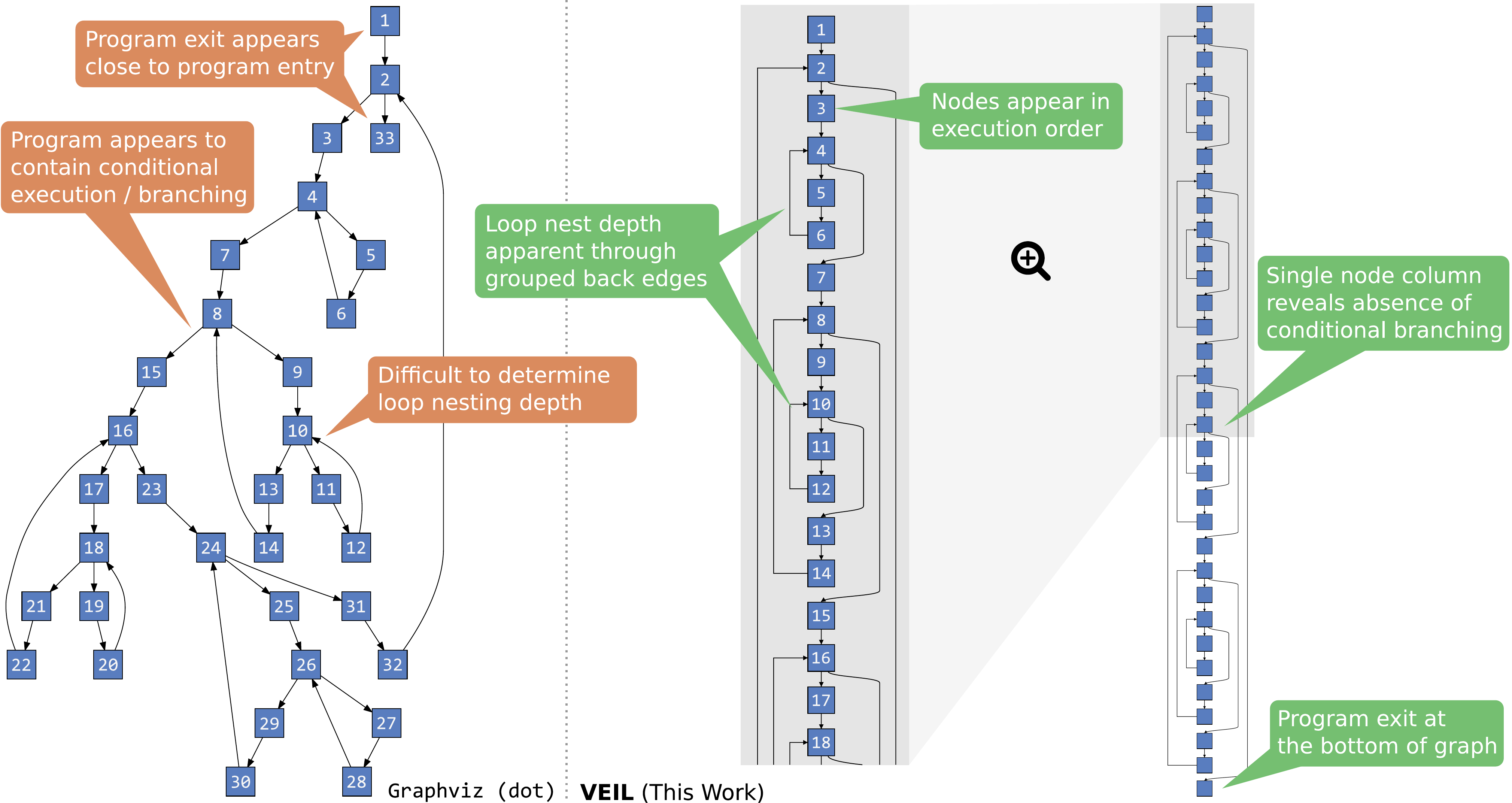}
    \caption{The control flow graph of a program computing 2-D electromagnetic wave propagation, drawn using Graphviz with the \texttt{dot} layout algorithm (left) and our new layout algorithm, VEIL (right), which optimizes layouts for control flow graphs.}
    \Description{The left side of the figure contains the control flow graph of a 2-D electromagnetic wave propagation, as drawn by Graphviz. In that drawing, the program exit node is placed close to (2 layers below) the program start, and various nodes contain two or more outgoing edges, which appears to form conditional branching in the code. Additionally, most loops are drawn in a way in which their exit / what is executed after the loop is drawn on the same height as or above the loop. This makes detecting loop nest depth difficult. The right side of the figure contains the same CFG laid out with VEIL. In this drawing, all nodes appear in a straight line from top to bottom, indicating that there is no conditional branching. The program exit lies at the very bottom of the graph, respecting happens-before relationships. Finally, back edges and forward edges are grouped together, respectively, more clearly indicating loop nest depths.}
    \label{fig:posterchild}
\end{teaserfigure}

\maketitle

\section{Introduction}
Control flow graphs (CFGs) are directed graphs that represent the order of execution in a program.
Visual analysis and understanding of CFGs are central to many tasks in computer systems, including debugging~\cite{devkota_cfgexplorer_2018, dux_visualizing_2005}, performance tuning~\cite{ferrante_program_1987, cummins_programl_2021, mansky_specifying_2016}, compiler development~\cite{devkota_ccnav_2021, hendren_visualization_2008}, reverse engineering~\cite{revelle_blaze_2023, goos_software_2002}, and security analysis~\cite{abadi_control-flow_2009, alam_annotated_2015}, to name a few.
In many of these contexts, CFGs are not just a convenient but necessary tool.
For instance, when source code is unavailable during binary analysis, or when compiler optimizations alter original control structures such as through simplification or reordering.
However, the CFGs of real-world applications are often very large, containing thousands of nodes and edges.
This scale makes visual comprehension difficult, especially in the presence of complex control flow constructs such as loops with multiple exit paths, extensive branching, or \texttt{goto} statements.

To support understanding and analysis, numerous \emph{graph drawing} algorithms have been developed to map graphs to the 2-D plane.
Considerable research has focused on defining graph layout criteria to maximize readability, often validated through comprehensive user studies~\cite{purchase_metrics_2002, purchase_empirical_2002, sugiyama_methods_1981, wong_evaluating_2006, burch_user_2017}.
This has led to the development of specialized graph drawing algorithms for specific classes of graphs, as well as some general-purpose algorithms that have become widely adopted in tools such as Graphviz~\cite{goos_graphviz_2002, gansner_technique_1993, gansner_drawing_2015}.

While general-purpose algorithms perform well across many domains, they often struggle with complex CFGs.
Their adherence to broad layout criteria can introduce artifacts that reduce readability.
For example, nodes that appear later in the execution order may be placed closer to the start of the program than nodes appearing earlier in execution order, disrupting the natural flow of analysis (see \autoref{fig:posterchild}).
Such artifacts suggest that state-of-the-art layout criteria are insufficient for the domain of CFGs and program flow analysis.

In this paper, we codify the criteria a graph drawing algorithm geared towards control flow graphs should follow.
We identify interaction and visualization techniques that benefit from the outlined criteria.
These criteria and techniques aim to help humans interact with CFGs in a more natural manner, more comparable to interactions with program source code rather than typical graph representations.

We develop a novel graph drawing algorithm called VEIL, the Vertical Execution-order Informed Layout, that optimizes CFG drawings with respect to the identified criteria.
By employing graph dominator analysis, our algorithm not only achieves intuitive graph layouts, but offers improved layout times compared to graph drawing algorithms frequently used for CFGs.

Using a series of CFGs from real-world applications, we perform an in-depth evaluation of our graph drawing algorithm and compare it to state-of-the-art techniques.
We evaluate and compare how well different techniques adhere to the identified layout criteria, and discuss how this impacts readability and comprehension.

In summary, this paper makes the following contributions:
\begin{itemize}
    \item A codified set of aesthetics criteria for evaluating CFG-specific graph layouts.
    \item VEIL, a novel layout algorithm that optimizes for CFG-specific aesthetics and structural criteria.
    \item An in-depth comparison of VEIL with state-of-the-art algorithms across real-world CFGs.
\end{itemize}

\section{Aesthetics of Control Flow Graph Drawings}\label{s:criteria}
Over the years, numerous studies have proposed concrete aesthetic criteria that graph drawing algorithms should follow to maximize readability and comprehension~\cite{bennett_aesthetics_2007, taylor_applying_2005, purchase_metrics_2002, purchase_empirical_2002, tamassia_automatic_1988}.
Many of these criteria have been repeatedly validated in empirical user studies~\cite{burch_state_2021, purchase_empirical_2002, huang_establishing_2013, huang_exploring_2010, purchase_experimental_1997, ware_cognitive_2002} and have become de facto standards for graph visualization.
By extension, algorithms for drawing control flow graphs (CFGs) should also adhere to these principles.
The nine most consistently supported criteria, i.e., those that appear frequently across studies and are rated as important for comprehension, are as follows:

\begin{figure}
    \centering
    \includegraphics[width=.9\linewidth]{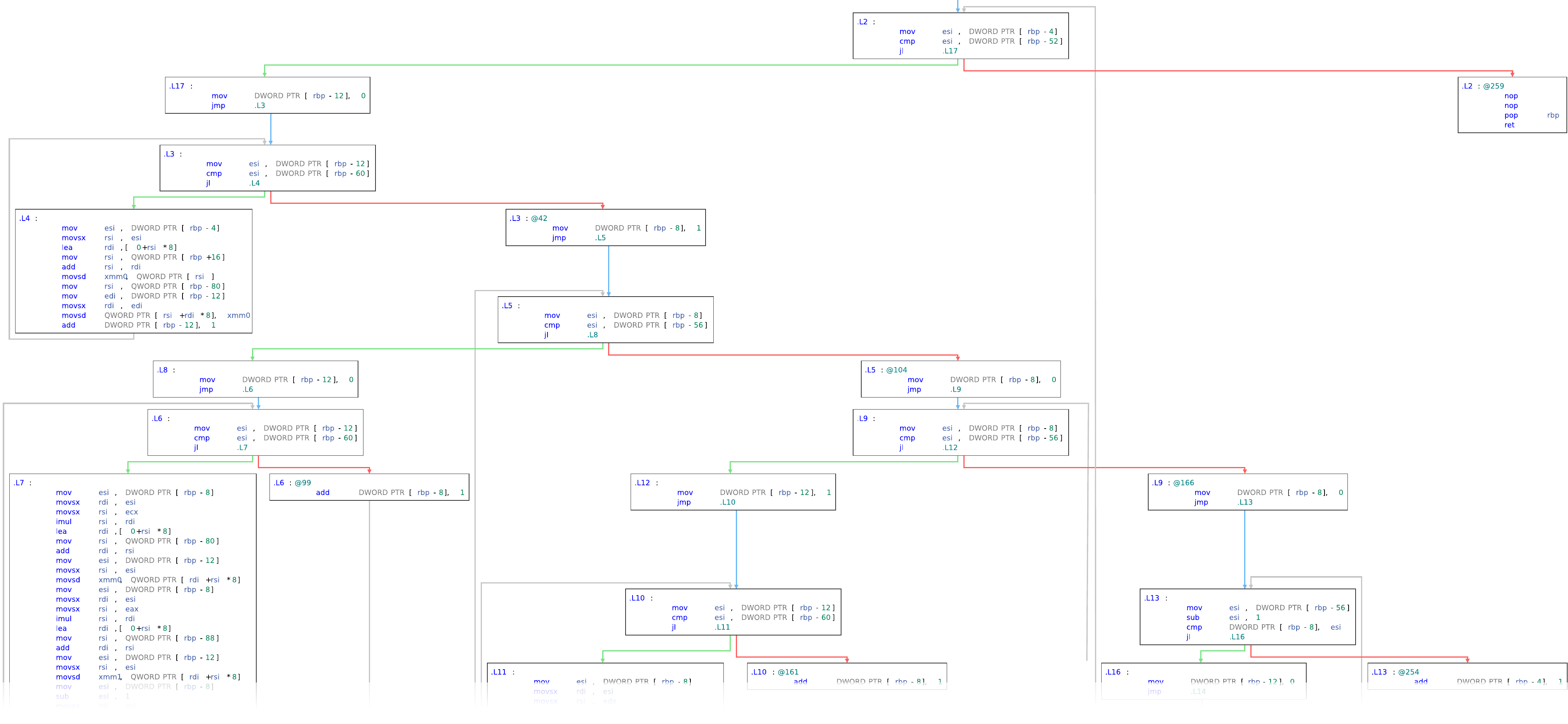}
    \vspace{-1.2em}
    \caption{Excerpt of a CFG as visualized by Compiler Explorer~\cite{godbolt}, showing x86 assembly instructions in each basic block.}
    \Description{The figure shows an excerpt of the control flow graph for the same program as in Fig. 1, drawn as created by Compiler Explorer. The control flow graph exhibits similar issues as the one from Fig. 1, laid out with Graphviz. However, in this figure, the nodes contain the actual x86 assembly instructions that are executed by each basic block, to demonstrate in what context CFGs are often visualized.}
    \label{fig:example-fdtd-godbolt}
\end{figure}

\begin{criterion}[Node Orthogonality]\label{c:node-orthogonality}
    Nodes should be evenly distributed on a virtual Cartesian grid without overlap, except when representing hierarchical nesting~\cite{purchase_metrics_2002, wong_evaluating_2006, huang_layout_2005, didimo_visualization_2018, purchase_empirical_2002}.
\end{criterion}

\begin{criterion}[Edge Orthogonality]\label{c:edge-orthogonality}
    Edges should run in straight horizontal or vertical lines~\cite{purchase_metrics_2002, wong_evaluating_2006, purchase_empirical_2002, burch_user_2017}.
\end{criterion}

\begin{criterion}[Edge Crossings]\label{c:edge-crossings}
    The number of edge crossings should be minimized~\cite{purchase_metrics_2002, tamassia_automatic_1988, purchase_empirical_2002, purchase_experimental_1997, huang_establishing_2013, huang_exploring_2010}.
\end{criterion}

\begin{criterion}[Edge Bends]\label{c:edge-bends}
    The number of edge bends, i.e., changes in direction along an edge, should be minimized~\cite{tamassia_automatic_1988, purchase_empirical_2002, purchase_experimental_1997, purchase_graph_2001, ware_cognitive_2002}.
\end{criterion}

\begin{criterion}[Edge Uniformity]\label{c:uniform-edge-length}
    Edge lengths should be kept uniform~\cite{taylor_applying_2005, davidson_drawing_1996, huang_aggregation-based_2012, huang_evaluating_2016}.
\end{criterion}

\begin{criterion}[Short Edges]\label{c:short-edges}
    The total edge length, as well as the maximum edge length, should be minimized~\cite{tamassia_automatic_1988, ware_cognitive_2002}
\end{criterion}

\begin{criterion}[Graph Area]\label{c:graph-area}
    The overall area occupied by the graph  should be minimized~\cite{tamassia_automatic_1988, taylor_applying_2005, wong_evaluating_2006, purchase_empirical_2002}
\end{criterion}

\begin{criterion}[Symmetry]\label{c:symmetry}
    Local and global symmetry should be maximized~\cite{welch_measuring_2017, biedl_perception_2018}
\end{criterion}

\begin{criterion}[Consistent Flow]\label{c:consistent-flow}
    In directed graphs, edges should follow a uniform flow direction~\cite{wong_evaluating_2006, purchase_metrics_2002}.
\end{criterion}

These criteria are not always mutually compatible, and tradeoffs are often necessary depending on the graph class.
Consequently, graph drawing algorithms typically optimize for a subset of criteria or adapt them to application-specific needs.
For example, prior work has investigated domain-specific aesthetics for object-relation diagrams (e.g., UML~\cite{UML})~\cite{purchase_graph_2001, wong_evaluating_2006, purchase_empirical_2002, eichelberger_aesthetics_2002} and for network graphs~\cite{gibson_survey_2013, misue_anchored_2008, wang_ambiguityvis_2016}.
By contrast, relatively little research has focused on improving layout quality specifically for control flow graphs beyond the general-purpose criteria above.

\begin{figure*}
    \centering
    \begin{subfigure}{.22\linewidth}
        \centering
        \includegraphics[width=.8\linewidth]{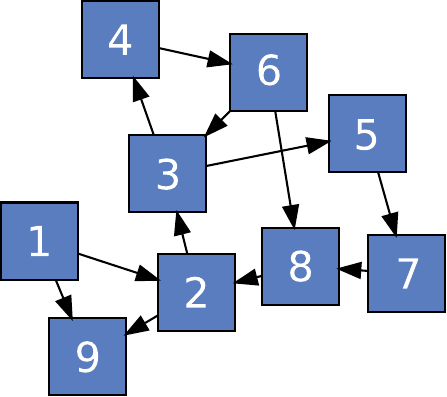}
        \caption{Force-directed placement, minimizing magnetic forces.}
        \Description{The CFG is laid out with force-directed placement, which minimizes magnetic forces. This leads to short edges, but at the cost of a lot of edge crossings and a violation of clear, uniform directionality. This makes detecting control flow constructs difficult.}
        \label{fig:fdp-demo}
    \end{subfigure}\hfill
    \begin{subfigure}{.22\linewidth}
        \centering
        \includegraphics[width=\linewidth]{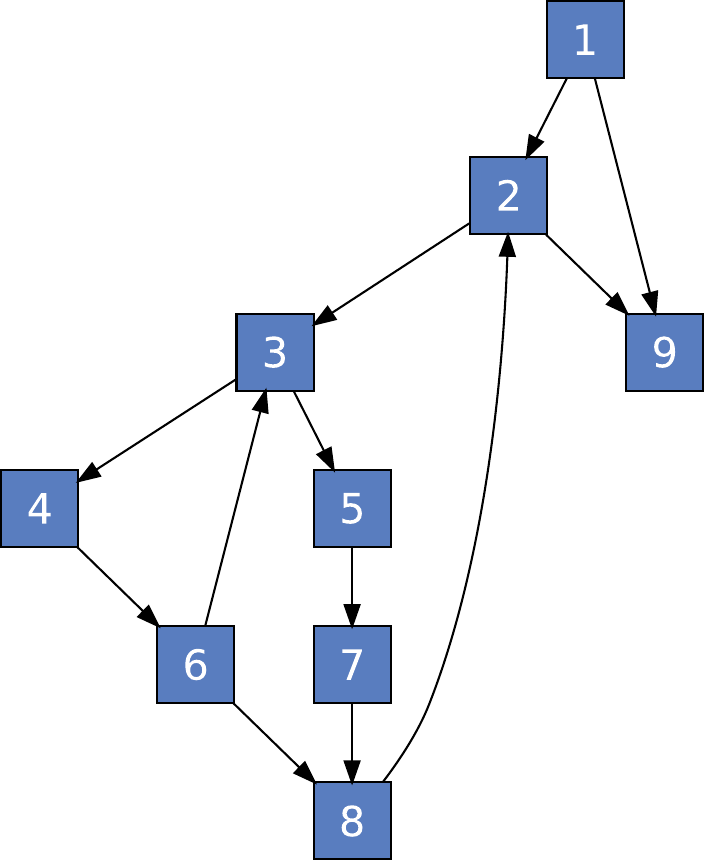}
        \caption{Layered graph drawing, optimizing for criteria \autoref{c:node-orthogonality}-\autoref{c:consistent-flow}.}
        \Description{The CFG is drawn using Graphviz, which respects most previously established, general graph aesthetic criteria well, and already improves readability.}
        \label{fig:dot-demo}
    \end{subfigure}\hfill
    \begin{subfigure}{.24\linewidth}
        \centering
        \includegraphics[width=.84\linewidth]{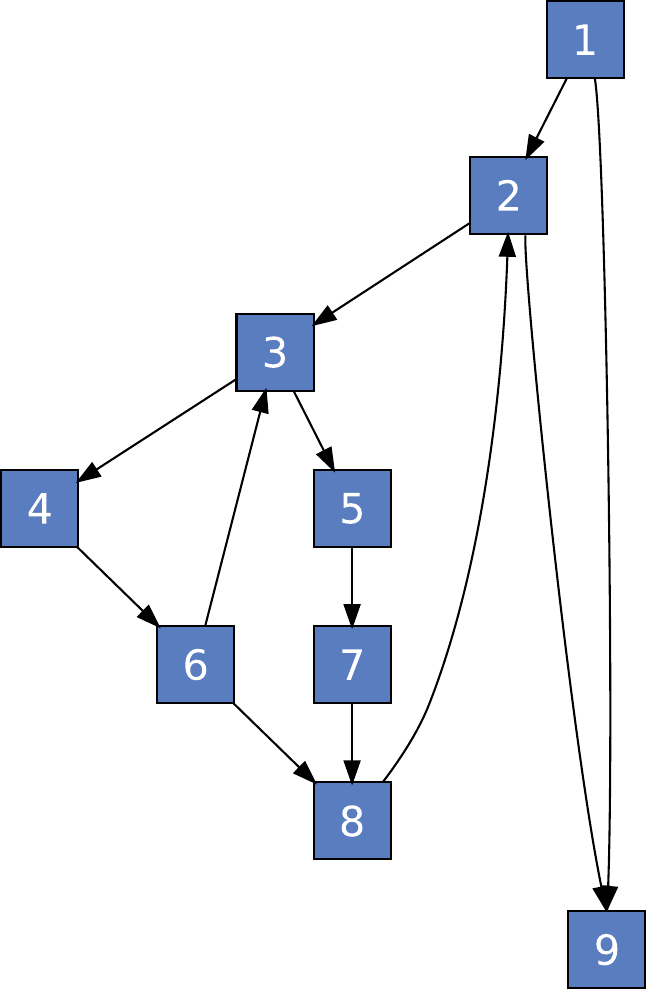}
        \caption{CFGExplorer~\cite{devkota_cfgexplorer_2018}, enforcing happens-before (\autoref{c:consistent-flow}) for loops.}
        \Description{The graph is drawn using CFGExplorer, which forces happens-before relationships for loops. This helps more clearly detect the presence of loops and somewhat improves readability.}
        \label{fig:cgexp-demo}
    \end{subfigure}\hfill
    \begin{subfigure}{.26\linewidth}
        \centering
        \includegraphics[width=.67\linewidth]{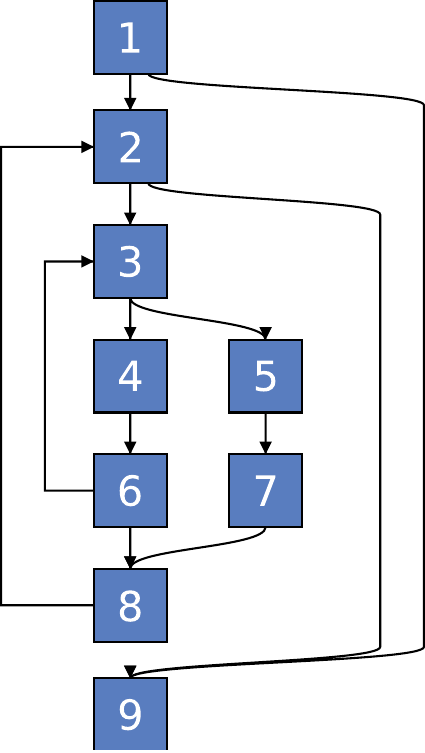}
        \caption{VEIL, enforcing edge direction grouping and all criteria \autoref{c:node-orthogonality}-\autoref{c:edge-grouping}.}
        \Description{Here, the graph is drawn using VEIL, which enforces happens-before relationships everywhere and enforces grouping of edges of the same type. Both improvements help detect where there is conditional branching, and how deep loop nests are.}
        \label{fig:veil-demo}
    \end{subfigure}
    \vspace{-1em}
    \caption{CFG of a conditionally executed 2-level loop nest drawn using different layouts to illustrate the effects of aesthetics criteria.}
    \Description{This figure contains a progression of control flow graphs from left to right, all of the same 2-level loop nest, demonstrating how the aesthetic criteria put forth by previous research and in this paper help in improving CFG readability.}
    \label{fig:demo-criteria}
\end{figure*}

\subsection{Control Flow Graphs and Aesthetics Criteria}
CFGs are a special class of directed graphs that capture the order in which operations in a program are executed.
A directed graph (digraph) is a pair $G = (V, E)$, where $V$ is a set of nodes and $E$ is a set of ordered pairs of nodes $e = (u, w)$, representing edges directed from $u$ to $w$.

CFGs form a subclass of digraphs with two additional requirements:
\begin{enumerate}
    \item A CFG has exactly one source node (a node with no incoming edges), representing the program entry.
    \item A CFG has at least one sink node (a node with no outgoing edges), representing program exit points.
\end{enumerate}

Each node in a CFG corresponds to a \emph{basic block}, which is a sequence of instructions with no branching into or out of the basic block.
Edges then represent possible control flow between blocks, encoding essential information about a program's structure and behavior.
For example, a node with multiple outgoing edges represents conditional branching (e.g., an \texttt{if-else} statement).
An example of a CFG can be seen in \autoref{fig:example-fdtd-godbolt}, which shows the instructions contained in each basic block and contains a number of edges forming loops and branching.
For simplicity, most of the CFGs shown in this paper will show basic blocks as abstract nodes with no contents (e.g., \autoref{fig:posterchild}, \autoref{fig:demo-criteria}), since the focus lies on the layout itself rather than visualization of instructions.

CFGs also represent a special case of digraphs in that they exhibit a few identifying characteristics.
Apart from programs with many large \texttt{switch-case} constructs, CFGs typically have a low average degree (few edges per node) and are thus sparse ($|E| \ll |V|(|V| - 1)$).
In addition to branching, CFGs usually contain \emph{back edges}, i.e., edges that form cycles~\cite{hecht_characterizations_1974}, which typically correspond to loop constructs such as \texttt{for} or \texttt{while} loops.
Both branching and back edges are critical for understanding program behavior, and detecting them reliably is essential for program analysis~\cite{devkota_ccnav_2021, revelle_blaze_2023}.

To illustrate how crucial established aesthetics criteria are, even when not adapted to CFGs, \autoref{fig:fdp-demo} shows the CFG of a conditionally executed 2-level loop nest, as laid out by force-directed placement using Graphviz's~\cite{goos_graphviz_2002} \texttt{fdp} algorithm.
Force-directed placement draws graphs by minimizing attractive and repulsive forces between nodes based on their distances to each other and whether they are connected through edges.
While this placement keeps edges short, consistent, and straight (\autoref{c:short-edges}, \autoref{c:uniform-edge-length}, \autoref{c:edge-bends}), and minimizes overall graph area (\autoref{c:graph-area}), it does not lead to edges pointing in a consistent direction (\autoref{c:consistent-flow}) or that avoid crossings (\autoref{c:edge-crossings}).
By violating both these criteria, readability of the drawing suffers compared to a drawing of the same CFG using a layered graph drawing algorithm (\texttt{dot}~\cite{gansner_drawing_2015}), shown in \autoref{fig:dot-demo}.

\subsection{Formalizing Control Flow-Specific Aesthetic Criteria}\label{ss:cfg-criteria}
Using the typical characteristics of CFGs, we formalize two additional control flow-specific aesthetic criteria for drawing CFGs to further improve readability, while still adhering to established graph drawing criteria.
When reasoning about execution order, developers are most familiar with source code, which presents instructions as a linear sequence within a text file.
Within a scope (e.g., a function or loop), the perceived execution order typically aligns with textual order, unless altered by compiler or runtime optimizations.
Any deviation, such as loops or branching, is explicitly marked with braces, indentations, or keywords, which provide visual cues for scope and control flow.

Given this strong mental model of \emph{top-to-bottom execution order} in source code, laying out a CFG in a similar manner should transitively improve readability and comprehension.
We formalize this intuition with the following control flow-specific aesthetic criterion:

\begin{criterion}[Happens-Before]\label{c:happens-before}
    For any two basic blocks $A$ and $B$, if $B$ is guaranteed to execute after $A$ (when both are executed), then $B$ should be drawn below $A$. Conversely, if $A$ must execute before $B$, then $A$ should appear above $B$.
\end{criterion}

This criterion ensures that any block executed after a loop is drawn below the loop body.
However, it may lead to longer edges between the loop header (the condition check) and the block immediately following the loop (the loop exit), which conflicts with edge uniformity (\autoref{c:uniform-edge-length}) and short edges (\autoref{c:short-edges}).
As a result, most existing algorithms do not enforce \autoref{c:happens-before}, as demonstrated in \autoref{fig:dot-demo}.

Despite this tradeoff, edge length can provide valuable structural cues \emph{if} \autoref{c:happens-before} is respected.
For example:
\begin{itemize}
    \item A long downward edge indicates a conditional branch that skips a large portion of the program, or, if it reaches the bottom, an early termination, with edge length proportional to the skipped portion of the program.
    \item A long upward edge (back edge) signals a loop, with its length correlating to loop size.
\end{itemize}
\autoref{fig:cgexp-demo} shows a drawing of the same CFG as \autoref{fig:dot-demo} but using CFGExplorer~\cite{devkota_cfgexplorer_2018}, which attempts to highlight loops, thus better adhering to \autoref{c:happens-before}.
In doing so, one such long downward edge that was less detectable in \autoref{fig:dot-demo} becomes more exposed, indicating a conditional branch skipping most of the program.

When edges are drawn orthogonally with minimal bends (\autoref{c:edge-orthogonality}, \autoref{c:edge-bends}), the negative impact of longer edges has further been shown through user studies to be significantly reduced~\cite{burch_state_2021, burch_user_2017}.
Thus, longer edges resulting from enforcing happens-before vertical ordering can enhance rather than hinder comprehension.

To strengthen this effect, we propose grouping semantically similar long edges.
Assuming \autoref{c:happens-before} is satisfied, three categories of long edges arise naturally:
\begin{enumerate}
    \item \textbf{Forward edges} that skip parts of the program in the form of conditional branches.
    \item \textbf{Back edges} that run upward, forming loops.
    \item \textbf{Loop-exit edges} connecting a loop header to the block immediately following the loop (loop exit).
\end{enumerate}
Routing all back edges on one side of the graph makes loops and their approximate size immediately visible.
Similarly, routing forward edges to the opposite side, grouped together, highlights conditionally executed code and early program termination.
Both classes of edges can be efficiently identified with a depth-first search in the graph~\cite{hecht_characterizations_1974}.
We capture this in a second control flow-specific aesthetic criterion:

\begin{criterion}[Edge Direction Grouping]\label{c:edge-grouping}
    Back edges (against program flow) should be grouped on one side of the graph, while forward edges should be grouped on the opposite side.
\end{criterion}
This highlights any violation of \autoref{c:uniform-edge-length} and \autoref{c:consistent-flow}, which carries semantic information in the context of CFGs.
A layout that respects both \autoref{c:happens-before} and \autoref{c:edge-grouping} is expected to be narrow and vertical when control flow is simple, but widen in regions with many conditional branches (e.g., large case distinctions).
This property makes structural complexity visually salient and facilitates rapid program understanding at a glance, while allowing graphs to be explored through vertical scrolling similar to text files, rather than panning.
Combining this with navigation features such as jumping to a given edge's start or end point allows for more efficient navigation of large CFGs.
An example of a layout respecting \autoref{c:happens-before} and \autoref{c:edge-grouping} can be seen in \autoref{fig:veil-demo}, where the CFG from \autoref{fig:cgexp-demo} is drawn using our new algorithm, VEIL.
By grouping back edges to the left, this layout more readily exposes the presence of a 2-level nested loop, and the drawing with \autoref{c:happens-before} indicates conditional branching through a deviation from a single-column layout towards the center.

\section{VEIL: Vertical Execution-Order Informed Layout}\label{s:algorithm}
The VEIL graph drawing algorithm falls into the category of layered graph drawing methods, also known as the Sugiyama Framework~\cite{sugiyama_methods_1981}.
Layered graph drawing arranges nodes of a directed graph into top-to-bottom layers (or ranks), ensuring that all edges, except back edges, point downward, thereby satisfying aesthetic criterion \autoref{c:consistent-flow} for consistent flow.
This well-studied framework forms the basis for graph drawing algorithms in many state-of-the-art tools, most notably Graphviz~\cite{goos_graphviz_2002} with the \texttt{dot} algorithm~\cite{gansner_drawing_2015, gansner_technique_1993}.

On a high level, layered graph drawing proceeds in three main phases:
\begin{enumerate}
    \item \textbf{Layer assignment}.
    Each node is assigned to a rank so that edges point from higher to lower layers.
    \item \textbf{Crossing minimization}.
    Nodes within each layer are reordered to reduce the number of edge crossings.
    Since finding an optimal layout with the fewest crossings is NP-complete~\cite{garey_crossing_1983}, a variety of heuristic methods have been developed to find good approximations~\cite{hutchison_fast_2005, goos_alternative_1997, gansner_technique_1993}.
    To allow for the use of such methods, edges that span multiple ranks are sub-divided into chains of single-rank edges through the help of virtual intermediate nodes.
    \item \textbf{Coordinate assignment}.
    Based on the layer assignments and node orderings, screen coordinates are assigned to all nodes and edges.
    Previously split up long edges spanning multiple layers are re-formed, removing virtual intermediate nodes (also referred to as \emph{dummy nodes}) and inserting edge bends in their stead.
\end{enumerate}

This procedure is particularly well suited to directed or hierarchical graph structures, such as CFGs, because it produces layouts that more effectively satisfy widely accepted graph aesthetics criteria (\autoref{c:node-orthogonality}-\autoref{c:consistent-flow}; see \autoref{s:criteria}) than alternative approaches.
Consequently, graph drawing algorithms following the Sugiyama Framework have become the de facto standard for visualizing CFGs~\cite{devkota_ccnav_2021, devkota_cfgexplorer_2018}, including for the widely used compiler toolchain Clang/LLVM~\cite{lattner_llvm_2004}.

However, the technique is inherently designed for directed acyclic graphs (DAGs), which is why the layer assignment phase typically begins by breaking cycles through back-edge reversal, only re-forming cycles during coordinate assignment.
Since CFGs often contain many cycles, this frequently produces layouts that violate the happens-before and edge grouping aesthetics criteria (\autoref{c:happens-before}/\autoref{c:edge-grouping}).
For example, in a cycle representing a \texttt{for}-loop, the control flow after the loop body is often placed on a higher rank than most of the loop, in an attempt to preserve short and uniform edge lengths (\autoref{c:uniform-edge-length}/\autoref{c:short-edges}; see \autoref{s:criteria}).
Because cycles are removed prior to ranking, the algorithm cannot recover the correct happens-before relationships once cycles are reintroduced.
Similarly, during crossing minimization, upward edges cannot be distinguished and therefore cannot be grouped together, making them less identifiable in the final layout.

To improve layouts for CFGs, VEIL leverages the three-phase structure of layered graph drawing, following the principle of orienting edges from higher to lower ranks, but introduces novel algorithms for the individual phases.
Specifically, both the layer assignment and crossing minimization phases are modified to account for back edges by leaving them intact.
This allows layer assignment to optimize rankings with respect to the happens-before criterion (\autoref{c:happens-before}) and enables in-layer permutations to group both back and forward edges (\autoref{c:edge-grouping}).
In addition, the coordinate assignment phase is refined to improve node and edge orthogonality (\autoref{c:node-orthogonality}/\autoref{c:edge-orthogonality}), minimize overall graph area (\autoref{c:graph-area}), and enhance symmetry (\autoref{c:symmetry}).
The following subsections describe each phase in turn.

\RestyleAlgo{ruled}
\LinesNumbered
\SetKwComment{Comment}{/* }{ */}
\begin{algorithm}[t]
\caption{VEIL layer assignment algorithm}\label{alg:layer-assignment}
\KwData{CFG $G$ with all node ranks initialized to $0$, empty queue $Q$}
\KwResult{CFG $G$ with all nodes assigned to their rank}
\If{number of sink nodes in G > 1}{
    Connect all sink nodes to a virtual sink node
}
$Q.push([G.start, 0])$\;
\While{$Q.length > 0$}{
  $[v, rank] \gets Q.pop()$\;
  $v.rank \gets \mathrm{max}(rank, v.rank)$\;
  \If{not $v.visited$}{
    $v.visited \gets true$\;
      $successors \gets \{w \in G.successors(v)\mid$ edge $(v, w)$ is not a back edge$\}$\;
      \uIf{there exists a back edge $(u, v)$ \& $\{w \in successors \mid w$ is not post-dominated by $u\} \neq \emptyset$ }{
        $[loopExit, exitRank] \gets handleLoop()$
        \Comment*[r]{Find loop exit node/rank}
        $Q.push([loopExit, exitRank])$\;
      }
      \ElseIf{$|successors| > 1$}{
      $[mergeNode, mergeRank] \gets handleBranch()$
      \Comment*[r]{Find node/rank where branch re-joins}
      $Q.push([mergeNode, mergeRank])$\;
      }
      \ForAll{$s \gets successors$}{
      $Q.push([s, rank + 1])$\;
      }
  }
}
$contractEmptyRanks()$
\Comment*[r]{Remove empty intermediate ranks}
\end{algorithm}

\subsection{Layer Assignment}
Layer assignment is a crucial phase in layered graph drawing, as it ensures that happens-before relationships are preserved.
VEIL's ranking algorithm enforces these relationships by performing a breadth-first traversal of the graph, beginning at the program entry node with an initial rank of $r_0 = 0$.
At each subsequent traversal step $i$, the rank is incremented according to $r_i = r_{i - 1} + \delta_r$, where $\delta_r$ is $1$ for most traversal steps.

When the traversal encounters the start of a loop, $\delta_r$ is increased sufficiently so that, by the time the traversal reaches the basic block immediately following the loop at step $k$, the assigned rank exceeds that of the final node in the loop at step $j$, i.e., $r_j \leq r_i + \delta_r$.
Conversely, when the traversal reaches a conditional split, the algorithm ensures that the rank of the merge point at step $k$, where the branches meet again, is greater than the maximum rank of all nodes in the conditional branches.

A high level overview of the algorithm is presented in \autoref{alg:layer-assignment}.
The handling of loops and conditional branching is discussed in detail in the following Sections~\ref{sss:veil-loops} and \ref{sss:veil-branch}.

\begin{figure*}
    \centering
    \begin{subfigure}{.45\linewidth}
        \centering
        \includegraphics[width=\linewidth]{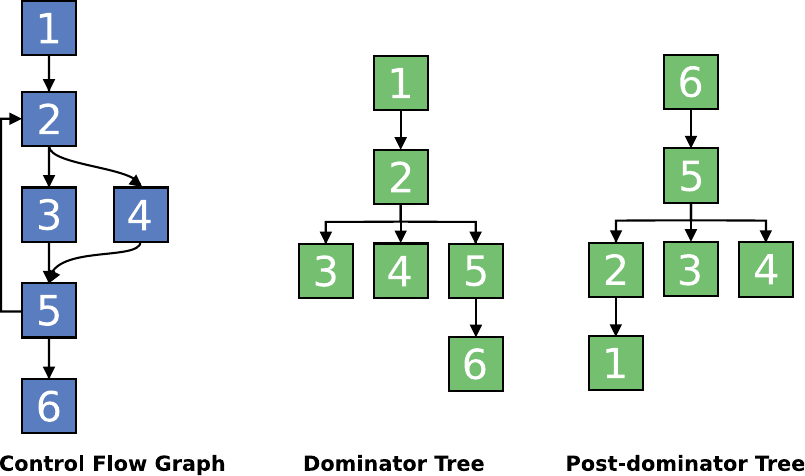}
        \caption{Inverted \texttt{do-while} style loop.}
        \Description{The figure shows an inverted loop together with its dominator and post-dominator trees.}
        \label{fig:inverted-loop}
    \end{subfigure}\hfill
    \begin{subfigure}{.45\linewidth}
        \centering
        \includegraphics[width=\linewidth]{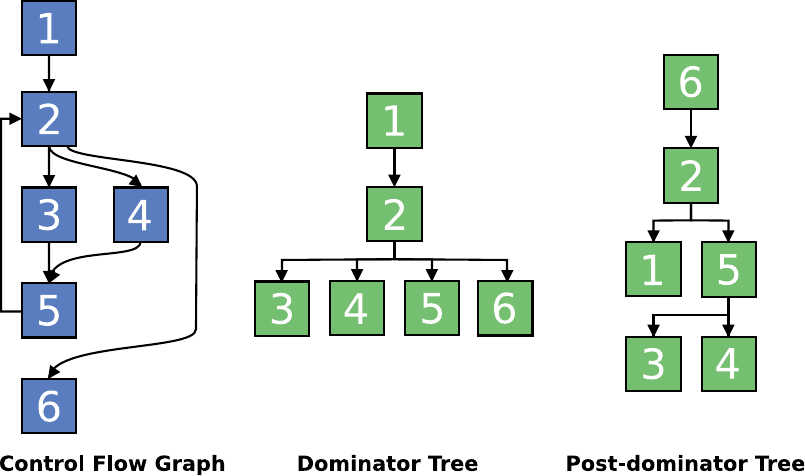}
        \caption{Regular \texttt{for/while} style loop.}
        \Description{The figure shows a regular loop together with its dominator and post-dominator trees.}
        \label{fig:reg-loop}
    \end{subfigure}
    \vspace{-1em}
    \caption{Examples of regular and inverted loops in CFGs together with their dominator and post-dominator trees.}
    \Description{The figure illustrates an inverted and a regular loop, respectively, and draws both dominator and post-dominator trees for both of them.}
    \label{fig:loops-dom-tree}
    \vspace{-1em}
\end{figure*}

\subsubsection{Handling Loops}\label{sss:veil-loops}
Control flow graphs may contain three types of loops, illustrated in \autoref{fig:loops-dom-tree}:
\begin{itemize}
    \item \textbf{Self-loops}. These consist of a single node with an edge pointing back to itself.
    No special handling is required in our algorithm (\autoref{alg:layer-assignment}), since a breadth-first traversal naturally assigns the exit rank one higher than that of the loop node.
    \item \textbf{Inverted loops (\texttt{do-while} style)}.
    In these loops, the condition is checked at the end of the body.
    Again, no special handling is needed: the exit is a successor of the condition check and will be traversed only after the loop body has been ranked.
    Cases involving branching or \texttt{break} statements are covered by the branching rules in \autoref{sss:veil-branch}.
    An example of an inverted loop's CFG sub-graph can be seen in \autoref{fig:inverted-loop}.
    \item \textbf{Regular loops (\texttt{for}/\texttt{while} style)}.
    Here, the condition is checked before the body, and special handling is required to correctly identify the loop exit and its rank.
    An example of this kind of CFG sub-graph can be seen in \autoref{fig:reg-loop}, where the condition check, also referred to as the \emph{loop header}, is seen in the node labelled 2.
\end{itemize}

To detect loops during traversal, we check for incoming back edges.
At traversal of node $v$, if any edge $(u, v)$ with $u \neq v$ is classified as a back edge, then $v$ is part of a loop:
either the start of the loop body in an inverted loop, or the loop header in a regular loop.
Back edges can be identified efficiently in $\mathcal{O}(|V| + |E|)$ through a preprocessing step using depth-first search (DFS)~\cite{hecht_characterizations_1974}.

To distinguish between inverted and regular loops, we examine whether successors of $v$ are post-dominated by $u$.
A node $x$ is \emph{dominated} by a node $y$, denoted $x \in \text{Dom}(y)$, if all paths from the program entry to $x$ pass through $y$.
Conversely, $y$ \emph{post-dominates} $x$, denoted $x \in \text{PDom}(y)$, if all paths from $x$ to the program exit pass through $y$.
The sets $\text{Dom}(y)$ and $\text{PDom}(y)$ consequently represent the set of nodes dominated and post-dominated by a node $y$, respectively.
Dominator and post-dominator trees, which are directed graphs where $x$ has an edge to $y$ if $x$ dominates or post-dominates $y$, respectively, can be precomputed efficiently in effectively linear time through a preprocessing step~\cite{cooper_simple_2006, sreedhar_incremental_1997}, and are sufficient for answering these queries.

\autoref{fig:inverted-loop} and \autoref{fig:reg-loop} show two examples of an inverted and regular control flow graph together with their corresponding control flow trees.
As can be seen in \autoref{fig:inverted-loop}, all successors of the node 2, which is the target of a back edge, are post-dominated by the source of the back edge, i.e., the node with label 5.
This identifies it as an inverted loop.
Conversely, in \autoref{fig:reg-loop} one successor (node 6) is \emph{not} post-dominated by node 5, and thus this identifies as a regular loop.

Once a regular loop headed at $v$ is detected, we identify the exit node $x$.
This node must be a successor of $v$, but it cannot be the back edge source $u$, which lies inside the loop body.
Successors of $v$ that dominate $u$ are also excluded, as they, too, lie inside the loop body.
If multiple candidates remain, such as when the loop head has several entry paths into the body, the candidate highest in the post-dominator tree (closest to the root) is chosen as the exit node $x$.

The rank of $x$ is then determined based on the size of the loop body, $|\text{loop}|$.
This is computed by counting nodes in a DFS from $v$, excluding back edges, forward edges, cross edge, and the exit edge $(v, x)$.
The exit rank is assigned as
\[
    x.rank = v.rank + |\text{loop}| + 1
\]
and traversal continues at each successor of $x$.

In loops with extensive branching, the calculation may result in $x.rank \gg v.rank + 1$, leaving empty intermediate layers.
These are eliminated in a cleanup pass at the end of the layer assignment phase, which contracts ranks to ensure that only occupied layers remain.

\subsubsection{Handling Conditional Splits}\label{sss:veil-branch}
A node $v$ visited in the traversal represents a conditional split if it has more than one successor (excluding back edges).
For such splits, the merge point can be determined by identifying the immediate post-dominator $w$ of $v$, i.e., the first successor $w$ of $v$ for which $w \in \text{PDom}(v)$.
For example, in the inverted loop in \autoref{fig:inverted-loop}, the loop body contains a conditional split starting at node 2.
A look at the post-dominator tree reveals that the first successor of node 2 which post-dominates, is node 5, which consequently represents the merge point for this split.

To ensure consistent happens-before rankings, the rank of $w$ must be at least one greater than the maximum rank obtained in any of branch of the split.
In \autoref{fig:inverted-loop}, this means that node 5 must be placed below both node 3 and node 4.
A direct way to compute this is to consider the difference between the number of nodes dominated by $v$ and those dominated by $w$.
This yields the total number of nodes contained within the conditional structure:
\[
|\text{cond}| = |\text{Dom}(v)| - |\text{Dom}(w)|
\]
The rank of w is then assigned as
\[
w.rank = v.rank + |\text{cond}| + 1
\]
As with loops, this calculation may overestimate the necessary distance between $v$ and $w$, leaving empty ranks, which are removed during cleanup at the end of the layer assignment phase.

\subsubsection{Time Complexity}
The time complexity of VEIL's layer assignment algorithm is governed by three key steps:
\begin{itemize}
    \item Ranking is performed in a graph traversal with complexity $\mathcal{O}(|V| + |E|)$.
    \item A pre-processing step needs to identify back edges, which can be done in $\mathcal{O}(|V| + |E|)$ time~\cite{hecht_characterizations_1974}.
    \item Dominator and post-dominator trees need to be computed, which has a worst-case complexity of $\mathcal{O}(|V| \cdot \mathcal{T}_d)$, where $\mathcal{T}_d$ is the dominator tree depth. However, in practice this typically takes linear time ($\mathcal{O}(|V|)$) for CFGs~\cite{cooper_simple_2006}.
\end{itemize}

By extension, ranking in VEIL has an asymptotic worst-time complexity of $\mathcal{O}(|V| \cdot \mathcal{T}_d)$, but is expected to complete in $\mathcal{O}(|V| + |E|)$ for most graphs.
Even in the worst case, this compares favorably with alternative layout algorithms, such as Dagre~\cite{dagre} or Graphviz's~\cite{goos_graphviz_2002} \texttt{dot}~\cite{gansner_drawing_2015}, where typical complexity is $\mathcal{O}(|E| \cdot |V| \cdot \log|V|)$.

\subsection{Crossing Minimization}
In standard layered graph drawing, crossing minimization begins by normalizing edges that span multiple ranks.
These edges are subdivided into chains of dummy nodes, reducing the problem to repeated bipartite crossing minimization between the nodes on consecutive ranks $V_l$ and $V_{l-1}$, with edges $E_{l}$ spanning the two layers.
Since finding the optimal ordering is NP-complete, heuristic methods are used~\cite{li_new_2002, goos_simple_2002, hutchison_fast_2005, eades_edge_1994}.
A widely adopted heuristic is the \textbf{barycenter method}, where each node $v \in V_l$ is positioned at the mean of the in-rank positions $u.ord$ of all its successors $u \in suc(v)$:
\[
v.ord = \frac{\sum_{w \in suc(v)}{w.ord}}{|suc(v)|}
\]
This heuristic is especially effective on sparse graphs~\cite{junger20022, makinen_experiments_1990}, which CFGs are typically a part of.
Crossing minimization is typically performed iteratively through alternating upward and downward sweeps of the various bipartite layer graphs until convergence or a cutoff is reached.

VEIL extends this process by keeping back edges intact and treating them explicitly during minimization.
When normalizing, back edges are assigned a distinct dummy node type, allowing them to be distinguished from forward edges during reordering.
Crossing minimization is then carried out in three steps:
\begin{enumerate}
    \item \textbf{Pre-sorting}.
    Each rank is ordered so that dummy nodes from normalized back edges appear first, followed by real nodes, and finally dummy nodes from forward edges. The horizontal position $v.ord$ of each node $v$ within each rank counts as the position for the purpose of the remaining steps.
    \item \textbf{Sweep with fixed back edges}.
    Upward and downward sweeps are performed to minimize crossings, but dummy nodes representing back edges remain fixed at the start of each rank.
    \item \textbf{Sweep with movable back edges}.
    A second sweep is performed where only dummy nodes for back edges are reordered, while all other nodes remain fixed.
\end{enumerate}
This design ensures that back edges are routed consistently to the left side of the layout, while forward edges are biased toward the right.
The result is a clearer grouping of long edges (\autoref{c:edge-grouping}; \autoref{ss:cfg-criteria}), improving the readability of CFGs.

\subsection{Coordinate Assignment}
The coordinate assignment phase of VEIL combines the rank assignments and the in-rank ordering information to distribute nodes on a Cartesian grid and then routes edges in straight lines between them.
User-configurable spacing values $\Delta_x$ and $\Delta_y$ define the horizontal and vertical distances between grid points.
With coordinates starting at the top left $(x = 0, y = 0)$, increasing rightward and downward, a node $v$ in rank $v.rank$ and with in-layer index $v.ord$ is placed as:
\[
(v.x, v.y) = (\Delta_x * (v.ord - |BE_{v.rank}|), \Delta_y * v.rank)
\]
where $|BE_{v.rank}|$ denotes the number of dummy nodes corresponding to back edges in that layer.
This placing forces back edges to be grouped together to the left of the graph, as per \autoref{c:edge-grouping}.
However, a user configuration can be used to swap the coordinate assignment function out for $(v.x, v.y) = (\Delta_x * v.ord, \Delta_y * v.rank)$, which instead creates ``indentations'' in the layout corresponding to loop nest depth, akin to source code.

Once initial positions are assigned, VEIL performs an \textbf{edge straightening} pass to align dummy nodes belonging to the same normalized edge on a common $x$-coordinate.
Given a set of dummy nodes $V_d$ representing one normalized edge, the common $x-$coordinate is computed as:
\[
x =
\begin{cases}
\min_{v \in V_d} v.x & \text{if the edge is a back edge},\\
\max_{v \in V_d} v.x & \text{otherwise}.
\end{cases}
\]
Edges are then de-normalized, with dummy nodes removed and bends reintroduced at their positions.
Optionally, a spline routing step can be applied to smooth out bends before finalizing the layout.

\section{Evaluation}\label{s:eval}
To demonstrate the effectiveness of VEIL for control flow graph drawing compared to state-of-the-art approaches, we have implemented VEIL as a layout plugin for the visualization component of the DaCe optimizing compiler~\cite{ben-nun_stateful_2019}.
Using this implementation, we compare the resulting layouts for real-world CFGs with those obtained through the default layout algorithm found in the tool, provided through a graph drawing library called Dagre~\cite{dagre}.
Dagre implements a version of layered graph drawing with the barycenter heuristic~\cite{junger20022, goos_simple_2002, goos_fast_2002} and is used by many web-based graph visualizations~\cite{devkota_cfgexplorer_2018, devkota_ccnav_2021, devkota_domain-centered_2022}.
We also compare against Graphviz's~\cite{goos_graphviz_2002} \texttt{dot}~\cite{gansner_drawing_2015} layout, which is one of the most widely used graph drawing algorithms for CFGs, serving as the default for CFGs extracted from Clang/LLVM~\cite{lattner_llvm_2004}.

\subsection{Measuring Aesthetics}\label{ss:measurements}
To quantitatively evaluate layouts against the established aesthetics criteria, we compute the following metrics:
\begin{itemize}
    \item[\textbf{\autoref{c:node-orthogonality}}] \textbf{Node Orthogonality}.
    Efficiency of distributing $|V|$ nodes across the Cartesian grid, scored $\in [0, 1]$:
    \[
    O_V = \frac{|V|}{(w + 1)(h + 1)}
    \]
    where $w$ and $h$ are the grid dimensions~\cite{purchase_metrics_2002}.
    \item[\textbf{\autoref{c:edge-orthogonality}}] \textbf{Edge Orthogonality}.
    Average alignment of edge segments $E_s$ with the $x$-axis, scored $\in [0, 1]$:
    \[
    O_{E_s} = 1 - \frac{1}{|E_s|}\sum_{e_s \in E_s}\frac{\min(\theta, |90 - \theta|, 180 - \theta)}{45}
    \]
    where $\theta$ is the angle of an edge segment to the $x$-axis~\cite{purchase_metrics_2002}.
    
    \item[\textbf{\autoref{c:edge-crossings}}] \textbf{Edge Crossings}. Total number of edge crossings per layout.
    
    \item[\textbf{\autoref{c:edge-bends}}] \textbf{Edge Bends}. Number of intermediate edge points that deviate from a straight line between surrounding points.
    
    \item[\textbf{\autoref{c:uniform-edge-length}}] \textbf{Edge Uniformity}. Median absolute deviation (MAD) of logarithmic edge lengths in pixels:
    \[
    \text{MAD} = \text{median}(|log(L_i) - log(\tilde{L})|)
    \]
    for the set of edge lengths $L_1$, $L_2$, ..., $L_{|E|}$, where $\tilde{L} = \text{median}(L)$.
    
    \item[\textbf{\autoref{c:short-edges}}] \textbf{Short Edges}. Total, maximum, and median edge lengths in pixels.

    \item[\textbf{\autoref{c:graph-area}}] \textbf{Graph Area}. Bounding box area of the graph in square pixels.
    
    \item[\textbf{\autoref{c:symmetry}}] \textbf{Symmetry}. Measured via the spring-force tension between nodes~\cite{fruchterman_graph_1991} (used in force-directed graph drawing algorithms, which inherently optimize for symmetry), reporting the sum and median of all per-node tensions.
        The per-node tension force $|\vec{F}_v|$ for a node $v$ is given by:
    \[
    |\vec{F}_v| = \Bigg|\underbrace{\sum_{w \in V}\left(\frac{\vec{\Delta}_{w}^{v}}{|\vec{\Delta}_{w}^{v}|} \cdot \frac{L_{U}^2}{log(|\vec{\Delta}_{w}^{v}|)}\right)}_{\text{repulsive forces}} + \underbrace{\sum_{w \in suc(v)}\left(\frac{\vec{\Delta}_{w}^{v}}{|\vec{\Delta}_{w}^{v}|} \cdot \frac{log(|\vec{\Delta}_{w}^{v}|)^2}{L_{U}}\right) - \sum_{w \in pred(v)}\left(\frac{\vec{\Delta}_{w}^{v}}{|\vec{\Delta}_{w}^{v}|} \cdot \frac{log(|\vec{\Delta}_{w}^{v}|)^2}{L_{U}}\right)}_{\text{attractive forces}}\Bigg|
    \]
    where $L_U$ is the ideal edge length, i.e., unit length, the distance in pixels between two ranks, and $\vec{\Delta}_{w}^{v}$ is the displacement vector between two nodes $v$ and $w$.
    
    \item[\textbf{\autoref{c:consistent-flow}}] \textbf{Consistent Flow}. Proportion of edge segments pointing from higher to lower ranks~\cite{purchase_metrics_2002}, i.e.:
    \[
        \mathcal{CF} = \frac{|\{(u, v) \in E \mid u.rank < v.rank\}|}{|E|}
    \]
    
    \item[\textbf{\autoref{c:happens-before}}] \textbf{Happens-Before}. Since exact evaluation requires topological sorting (computationally infeasible for cyclic graphs), we approximate by measuring the relative rank of the program exit (sink) node to the maximum rank in the graph:
    \[
    \mathcal{HB} = \frac{exit.rank}{|ranks|}
    \]
    
    \item[\textbf{\autoref{c:edge-grouping}}] \textbf{Edge Direction Grouping}. Median of the minimal distances between any two pairs of back edges or forward edges with overlapping $y$-coordinates, respectively.
\end{itemize}
For the metrics measuring \autoref{c:node-orthogonality}, \autoref{c:edge-orthogonality}, \autoref{c:consistent-flow}, and \autoref{c:happens-before}, higher values are better; for \autoref{c:edge-crossings}--\autoref{c:symmetry} and \autoref{c:edge-grouping}, lower values are better.

\begin{figure}
    \centering
    \includegraphics[width=\linewidth]{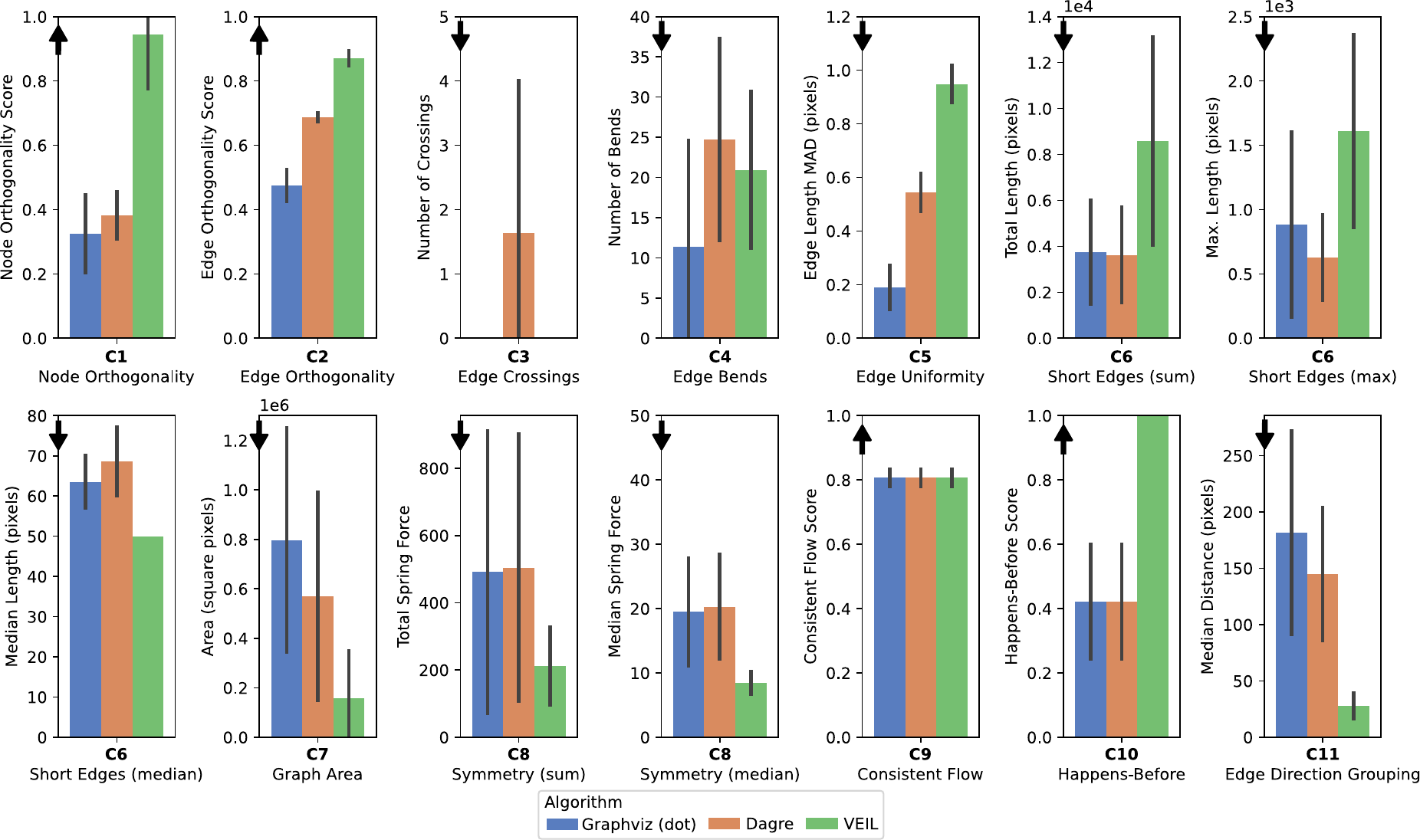}
    \vspace{-2em}
    \caption{Criteria metrics across all 30 Polybench applications. Arrows indicate whether higher ($\uparrow$) or lower ($\downarrow$) values are better.}
    \Description{The figure shows 14 plots, each of which comparing a particular graph aesthetic metric between Graphviz, Dagre, and VEIL, as measured across all 30 Polybench applications. The plots illustrate clear improvements in layout performance for VEIL across all metrics, with the exception of overall edge uniformity and length. However, as discussed in the text, the longer edges convey clear semantic information in VEIL, thus helping readability.}
    \label{fig:polybench-all}
\end{figure}

\subsection{Polybench}\label{ss:polybench}
The Polybench/C~\cite{polybench} benchmark suite provides 30 applications representative of program patterns commonly encountered during performance optimization.
We use these applications to extract 30 CFGs that are representative of the types of graphs encountered when working with CFGs for compiler development or performance tuning.
Each graph is drawn using VEIL, Dagre~\cite{dagre}, and Graphviz's \texttt{dot}~\cite{goos_graphviz_2002, gansner_drawing_2015}.
For each layout, we compute the metrics described in \autoref{ss:measurements}.
\autoref{fig:polybench-all} summarizes the results across  all 30 CFGs, reporting median values with standard deviation error bars.

For node and edge orthogonality (\autoref{c:node-orthogonality}, \autoref{c:edge-orthogonality}), VEIL achieves significantly higher scores than both Dagre and \texttt{dot}.
In edge crossings (\autoref{c:edge-crossings}), VEIL and \texttt{dot} consistently avoid crossings, while Dagre does not.
The number of edge bends (\autoref{c:edge-bends}) is comparable across all algorithms.
However, in VEIL, bends correspond directly to meaningful control-flow constructs, such as loops or conditional skips.

As expected from enforcing happens-before relationships, VEIL exhibits higher edge length variability (\autoref{c:uniform-edge-length}) and longer total and maximum edge length (\autoref{c:short-edges}) compared to the baselines.
Yet, VEIL achieves a lower median edge length with zero variance, indicating that the majority of edges are unit-length (spanning a single rank).
Similar to edge bends, longer edges in VEIL represent semantically important control flow constructs.

VEIL also outperforms the baselines on graph area (\autoref{c:graph-area}) and symmetry (\autoref{c:symmetry}), reducing area by up to \textbf{5}$\times$ and global spring force by $2.4\times$ compared to the baselines.
Consistent flow (\autoref{c:consistent-flow}) is identical across all three algorithms, while VEIL uniquely guarantees a perfect happens-before score (\autoref{c:happens-before}), representing a $2.4\times$ improvement over Dagre and \texttt{dot}.
Additionally, similar edge types in VEIL are more tightly grouped (\autoref{c:edge-grouping}), leading to an up to $7.4\times$ reduction in the distance between similar edge types compared to the baselines.

The improved algorithmic complexity of VEIL's layer assignment phase compared to Dagre and \texttt{dot} additionally leads to an average of $1.4\times$ faster layout times than measured in Dagre.

\begin{figure*}
    \centering
    \begin{subfigure}{.21\linewidth}
        \centering
        \includegraphics[width=.7\linewidth]{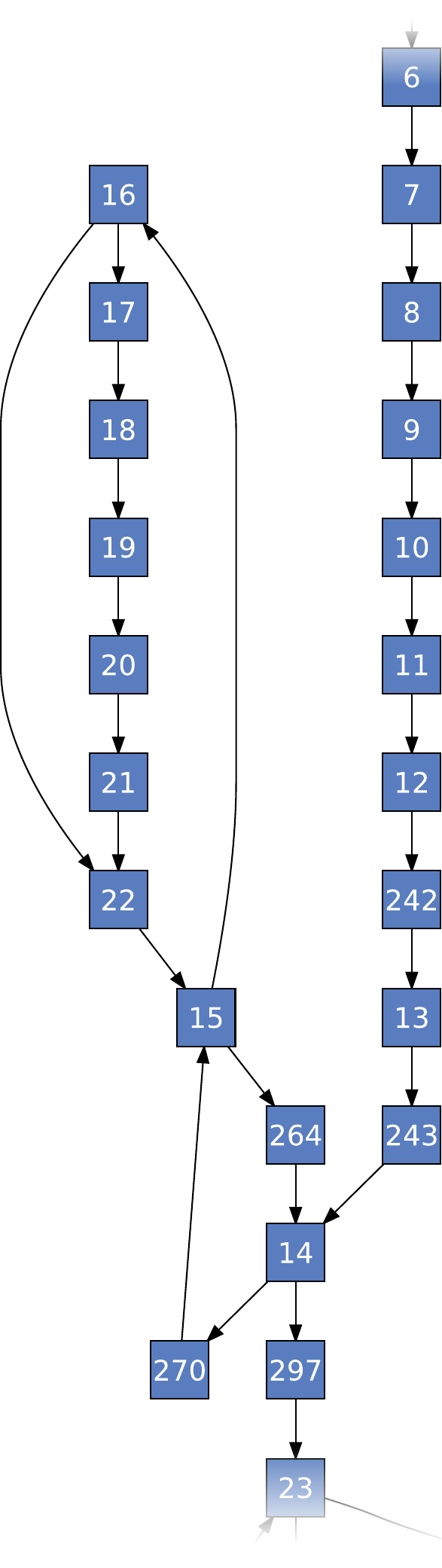}
        \caption{First loop nest (\texttt{dot})}
        \Description{The first loop nest in the CFG of CLOUDSC is drawn using Graphviz. In this drawing, the loop nest awkwardly runs upwards alongside a sequence of other control flow blocks, making it difficult to interpret the loop nest.}
        \label{fig:cloudsc-graphviz-backwards-loop}
    \end{subfigure}\hfill
    \begin{subfigure}{.21\linewidth}
        \centering
        \includegraphics[width=.78\linewidth]{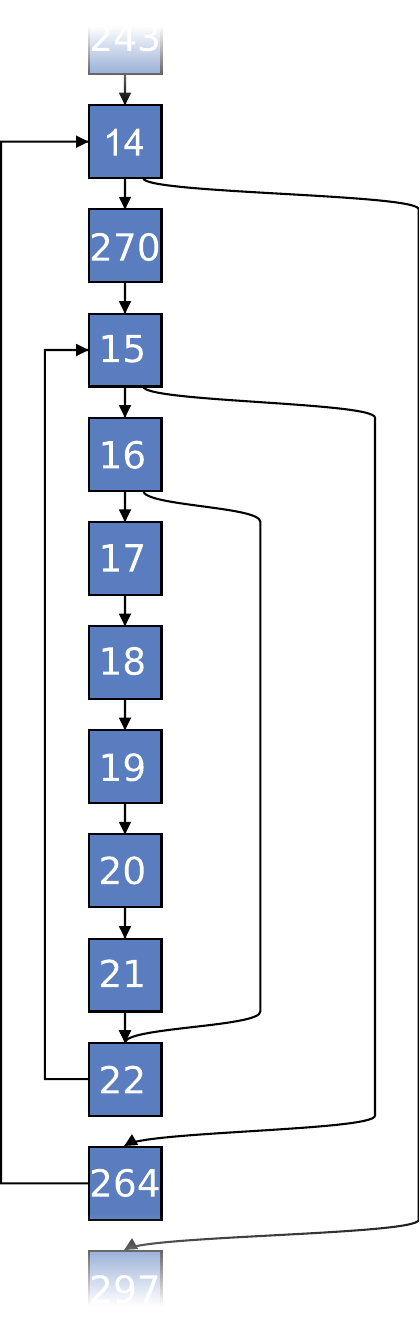}
        \caption{First loop nest (VEIL)}
        \Description{The same first loop nest is laid out with VEIL, which correctly enforces happens-before relationships and groups the back edges together. This makes it clear that it is a 2-level loop nest with no conditional branching.}
        \label{fig:cloudsc-backwards-loop-veil}
    \end{subfigure}\hfill
    \begin{subfigure}{.28\linewidth}
        \centering
        \includegraphics[width=\linewidth]{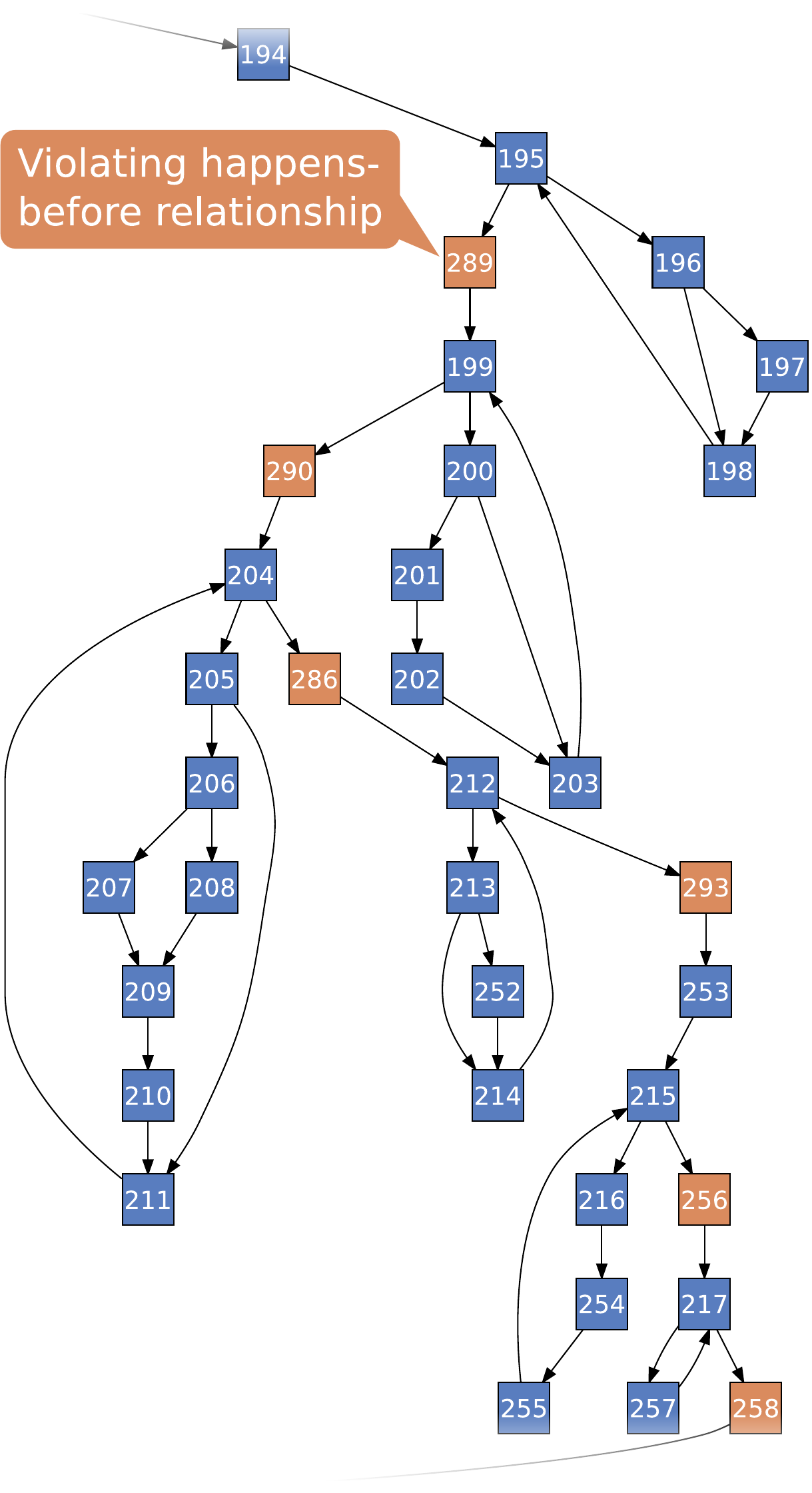}
        \caption{\texttt{dot}: Orange nodes violate \autoref{c:happens-before}.}
        \Description{A later loop nest in the CFG is laid out with Graphviz, where multiple loop exit nodes violate the happens-before relationships. As such, detecting the loop orders and nesting depths is difficult.}
        \label{fig:cloudsc-graphviz-loop-exits}
    \end{subfigure}\hfill
    \begin{subfigure}{.25\linewidth}
        \centering
        \includegraphics[width=.6\linewidth]{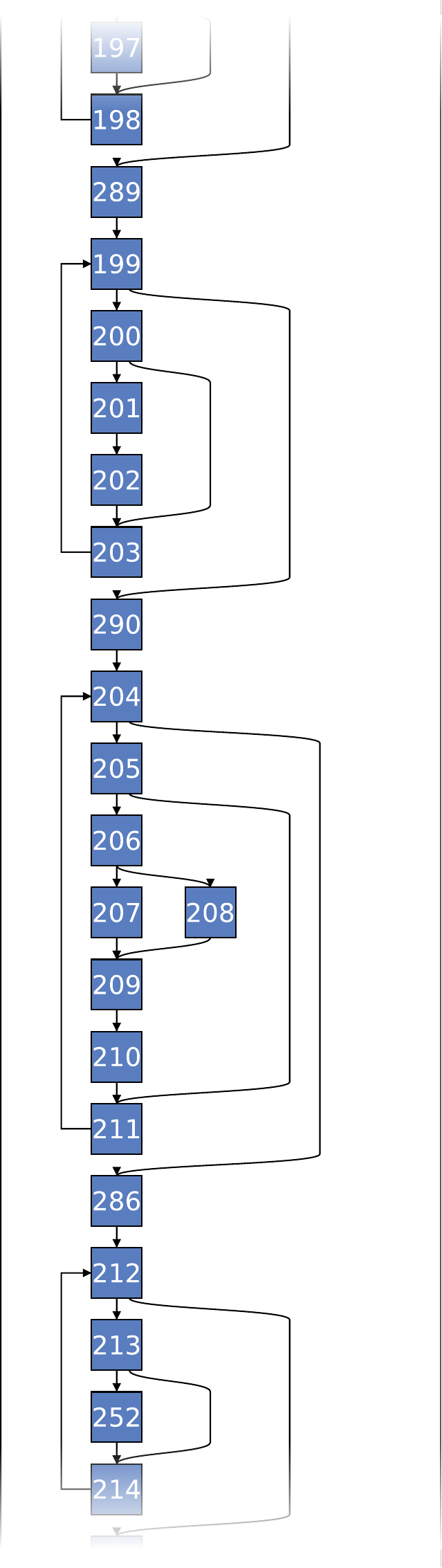}
        \caption{VEIL: No \autoref{c:happens-before} violations.}
        \Description{The same loop nests are drawn using VEIL, where edge grouping and clear happens-before layouts again help detect the concrete loop sequence and nesting depth. Additionally, the presence of a further back edge to the very left indicates that the entire loop sequence is nested inside an additional, bigger loop -- a fact that is not visible in the Graphviz drawing.}
        \label{fig:cloudsc-veil-loop-exits}
    \end{subfigure}
    \vspace{-1em}
    \caption{Various loop nests found in CLOUDSC demonstrate how existing approaches (\texttt{dot}) fail to respect happens-before relationships and edge direction grouping, decreasing readability. VEIL drawings of the same loop nests offer improved readability.}
    \Description{The figure shows two excerpts of the CLOUDSC CFG, as laid out with Graphviz and VEIL, respectively.}
    \label{fig:cloudsc-artifacts}
\end{figure*}

\subsection{CLOUDSC}
The European Centre for Medium-Range Weather Forecasts' (ECMWF) Integrated Forecasting System (IFS) contains a physics simulation, the cloud microphysics scheme (CLOUDSC)~\cite{cloudsc}, which exhibits complex control flow patterns typical for such simulations.
This class of applications frequently features deep nested loops, state-dependent branching (i.e,. large, different code paths), early termination, or error correction.
Such patterns often produce layouts that violate happens-before relationships, making CFGs difficult to interpret.

\autoref{fig:cloudsc-graphviz-backwards-loop} illustrates this problem in a CFG section drawn with Graphviz/\texttt{dot}.
After an initial linear sequence of nodes, the flow unexpectedly turns upwards, introducing two consecutive upward edges.
While loops are present, the upward edges do not correspond to back edges but to loop bodies, obscuring their structure.
In contrast, VEIL (\autoref{fig:cloudsc-backwards-loop-veil}) correctly identifies loop bodies and exits using dominator analysis, arranges nodes in proper topological order, and routes only back edges upwards.
Edge direction grouping (\autoref{c:edge-grouping}) further improves readability by keeping back edge consistently on the left, revealing loop nesting, while skip edges are kept right to highlight the conditional execution of the entire loop body.

A similar issue is shown in \autoref{fig:cloudsc-graphviz-loop-exits}, where loop exit nodes appear above their corresponding loop bodies in the \texttt{dot} layout, violating happens-before and hindering loop identification.
VEIL (\autoref{fig:cloudsc-veil-loop-exits}) places exit nodes below their bodies and groups back edges on the left, making nested loops and their extents immediately visible.
The presence of a second, long back edge on the far left even exposes the existence of a larger enclosing loop, which is not discernible in the \texttt{dot} layout.

\begin{figure*}
    \centering
    \begin{subfigure}{.6\linewidth}
        \centering
        \includegraphics[width=\linewidth]{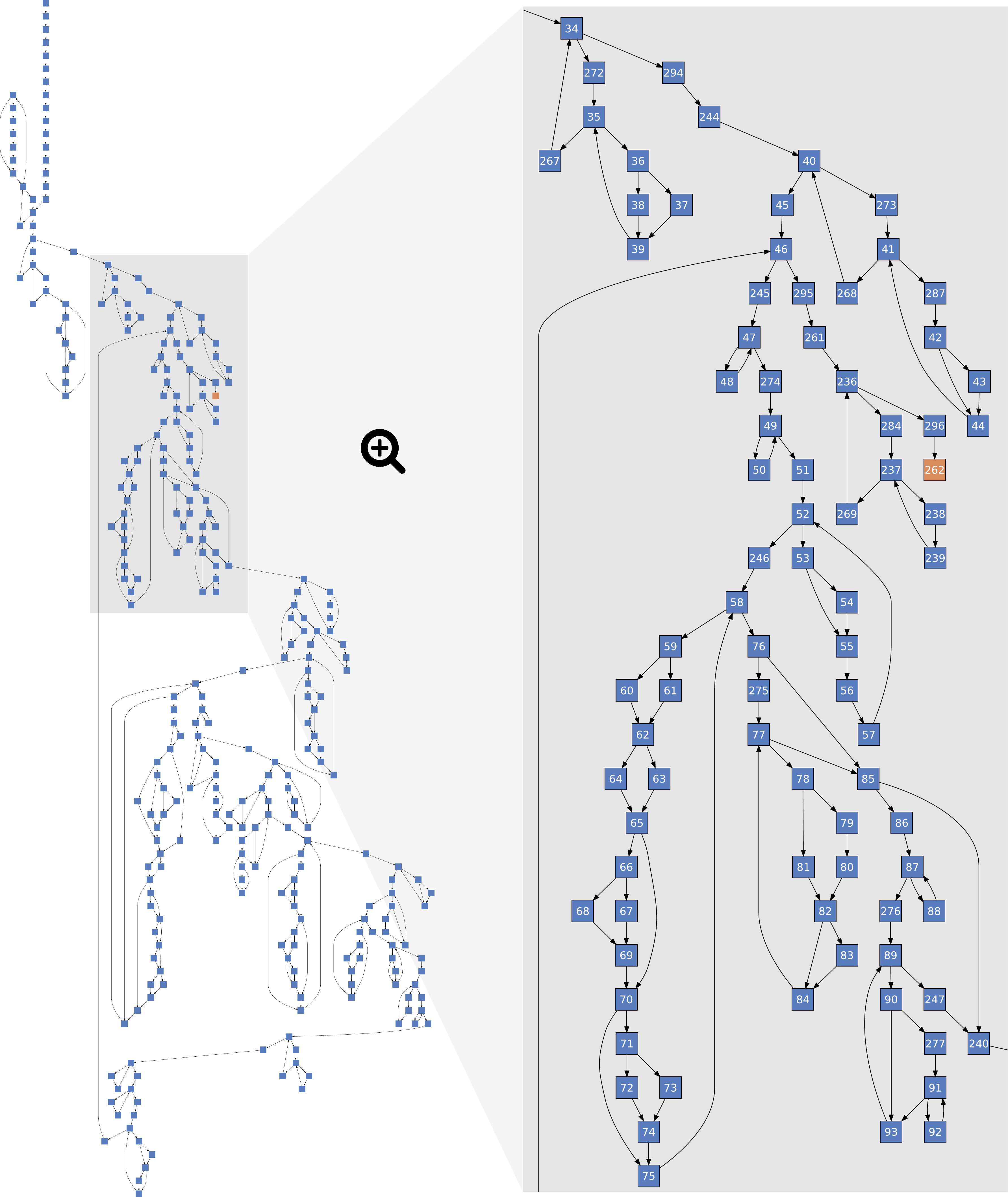}
        \caption{Graphviz's \texttt{dot} places the program exit (orange) close to the top of the graph.}
        \Description{The figure shows the full CFG drawn with Graphviz, including a zoomed in section of the sub-graph surrounding the program exit. In the figure it is visible that the program exit is placed roughly 2/3 up the height of the drawing, clearly violating happens-before relationships.}
        \label{fig:cloudsc-graphviz-exit-rank}
    \end{subfigure}\hfill
    \begin{subfigure}{.39\linewidth}
        \centering
        \includegraphics[width=.63\linewidth]{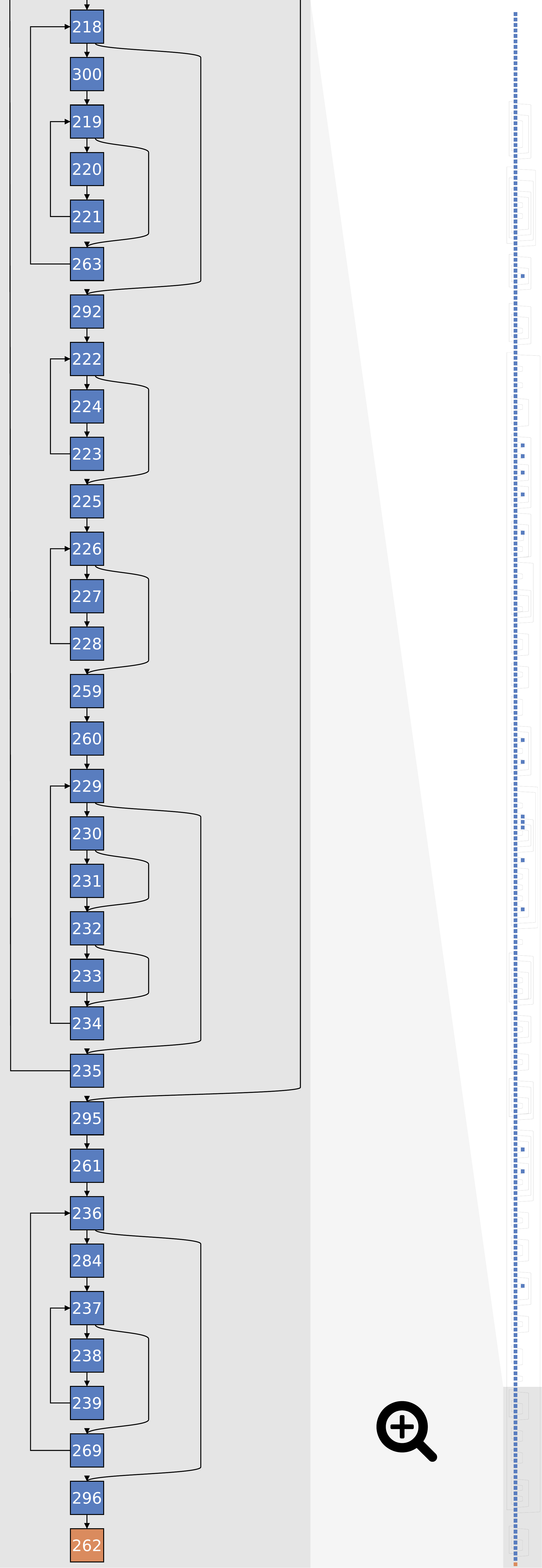}
        \caption{VEIL keeps the program exit at the bottom.}
        \Description{The figure shows the full CFG drawn with VEIL, including the same zoomed in section around the program exit. Here, the program exit is drawn well at the bottom of the graph.}
        \label{fig:cloudsc-exit-rank-veil}
    \end{subfigure}
    \vspace{-1em}
    \caption{Established layouts violate happens-before relationships (\autoref{c:happens-before}) by drawing CLOUDSC's program exit (orange) close to the top.}
    \Description{The figure shows the full CLOUDSC CFG laid out with Graphviz and VEIL, respectively, including zoomed in sections of the CFG around the program exit.}
    \label{fig:cloudsc-artifacts-2}
\end{figure*}

\autoref{fig:cloudsc-graphviz-exit-rank} shows another example, where \texttt{dot} places the program exit block in the upper third of the layout, surrounded by other nodes.
Locating the program exit and early terminations leading to it thus requires manually scanning for sink nodes, a task that becomes infeasible in graphs with thousands of nodes.
VEIL (\autoref{fig:cloudsc-exit-rank-veil}) instead places the exit at the bottom of the layout, making program termination trivial to locate and clearly showing a final nested loop which each program exit path runs through.

\begin{figure*}
    \centering
    \begin{subfigure}{.5\linewidth}
        \centering
        \includegraphics[width=.25\linewidth]{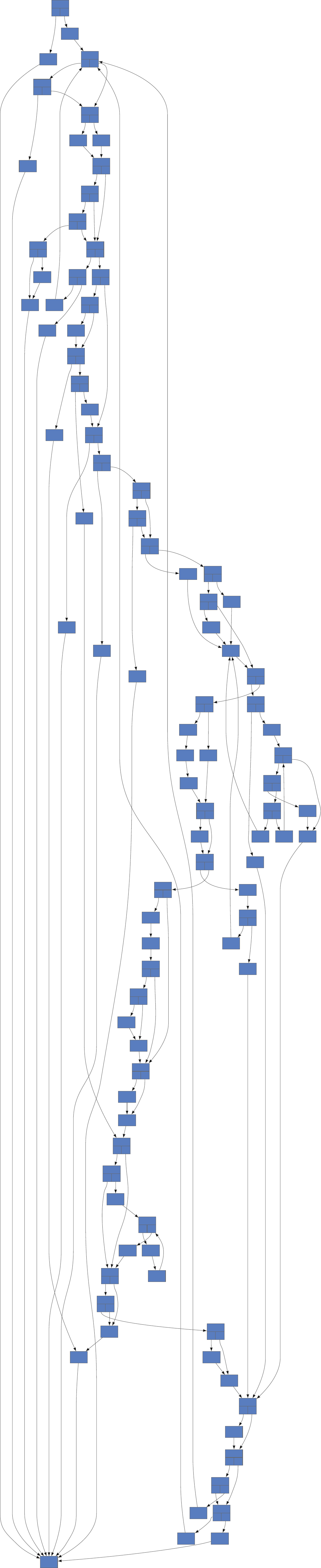}
        \caption{Graphviz's layout leads to mixing of back edges and forward edges, making control flow paths formed by \texttt{goto}s hard to trace.}
        \Description{The full CFG drawn with Graphviz is shown, which demonstrates that the mixing of back edges and forward edges resulting from irregular control flow expressed with goto statements makes analyzing the sequence of operations and possible execution paths difficult.}
        \label{fig:shmem-graphviz}
    \end{subfigure}\hfill
    \begin{subfigure}{.46\linewidth}
        \centering
        \includegraphics[width=.25\linewidth]{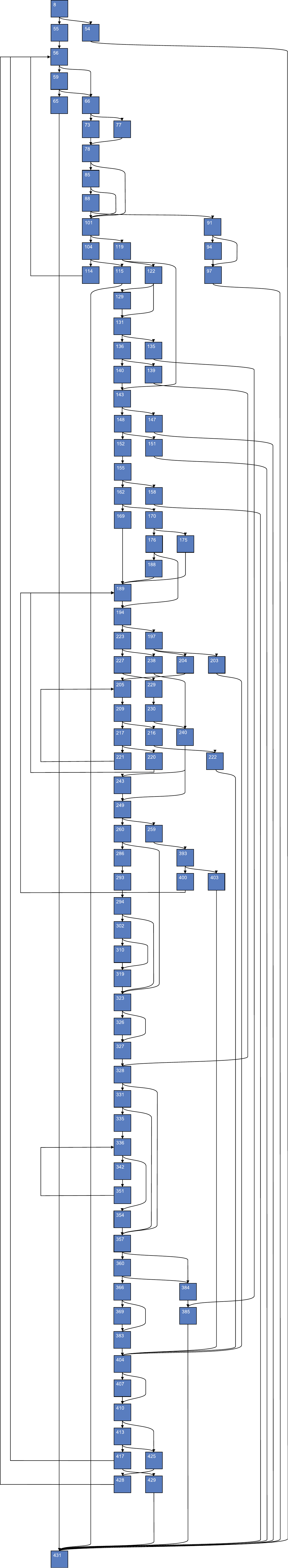}
        \caption{Through VEIL's edge direction grouping (\autoref{c:edge-grouping}), the quasi-loop structures formed by \texttt{goto}s are exposed.}
        \Description{The same CFG is drawn with VEIL, where the grouping of back edges and forward edge helps in detecting how goto statements form quasi-loops, which helps in analyzing possible code execution paths, improving security analysis and debugging.}
        \label{fig:shmem-veil}
    \end{subfigure}
    \vspace{-1em}
    \caption{The CFG of Linux's \texttt{shmem\_getpage\_gfp} memory page allocation function, drawn via Graphviz and VEIL.}
    \Description{The full CFG of the memory page allocation function is drawn using Graphviz and VEIL, respectively.}
    \label{fig:shmem-case}
\end{figure*}

\subsubsection{Quantitative Comparison}
Quantitatively, VEIL achieves up to \textbf{10}$\times$ higher node orthogonality (\autoref{c:node-orthogonality}) and $1.7\times$ higher edge orthogonality (\autoref{c:edge-orthogonality}) than Dagre and \texttt{dot}, along with a $2.5\times$ reduction in edge bends (\autoref{c:edge-bends}).
While \texttt{dot} produces slightly more uniform and shorter edges (\autoref{c:uniform-edge-length}, \autoref{c:short-edges}), VEIL emphasizes unit-length edges and leverages longer edges to highlight loops and skips (cf. \autoref{fig:cloudsc-backwards-loop-veil}).
Median edge length in VEIL is up to $1.8\times$ shorter than in the baselines, underscoring its bias toward compact local flow.
VEIL also consistently achieves a happens-before score (\autoref{c:happens-before}) of 1, representing a $3\times$ improvement over the baselines for CLOUDSC.
Thanks to the reduced ranking time complexity, VEIL additionally completes the layout $1.6\times$ faster than Dagre.
Finally, the resulting slender, vertical layout reduces overall graph area (\autoref{c:graph-area}) by up to \textbf{13.1}$\times$.

\subsection{Linux Kernel Memory Management}
To demonstrate how VEIL helps to interpret irreducible / irregular control flow, we analyze the \texttt{shmem\_getpage\_gfp} function inside the Linux~\cite{linux} operating system.
The function is designed to find a memory page in cache or swap, or allocate it if can not be found.
As with much of the Linux kernel code, the function is characterized by a number of \texttt{goto} statements (12 \texttt{goto} statements and 6 labels) which form unstructured / irreducible control flow.
The function also contains some conditional branching, early return statements, and loops.
As such, even the code can be difficult to analyze, and the resulting CFG layout in Graphviz's \texttt{dot} becomes difficult to read, with edge crossings, a large number of upwards running edges, and many long, ungrouped edges (\autoref{fig:shmem-graphviz}).

Through VEIL's grouping of back edges, control flow becomes more readable and the back edges indicate how \texttt{goto} statements in the loop serve to form quasi-loops.
Similarly, various long skip edges routed to the right indicate multiple early return paths, while illustrating how \texttt{goto} statements are used to run error handling before exiting.
Analysis of codes containing such irregular control flow constructs is difficult even in code, which makes both debugging and security vulnerability analysis of such codes challenging.
Consequently, edge grouping helps for a mental map by more clearly showing repeat or skipping behavior, helping in tracing possible execution paths through the program.

\section{Related Work}
Graph drawing and software visualization are both highly active research areas.
We discuss a few of the advances and techniques most relevant or closely related to the contributions made in this paper, and discuss how VEIL separates itself from them.

\paragraph{General Graph Drawing}
Numerous general graph drawing algorithms have been developed to improve the readability of complex and large graphs.
The Sugiyama Method~\cite{sugiyama_methods_1981}, or layered graph drawing, has become the most popular framework for drawing directed acyclic graphs, due to the readability benefits obtained by its strict enforcement of a consistent flow direction.
Many state-of-the art algorithms build on that framework, most notably Graphviz~\cite{goos_graphviz_2002} using the \texttt{dot}~\cite{gansner_drawing_2015} algorithm, the VCG tool~\cite{goos_graph_1995}, or Dagre~\cite{dagre}.
Various improvements have been made to different parts of layered graph drawing, including novel heuristic methods for reducing edge crossings~\cite{junger20022, goos_alternative_1997, eades_edge_1994}.

Dig-CoLa~\cite{dwyer_dig-cola_2005, dwyer_drawing_2006} combines layered drawing with force directed layout techniques to improve edge direction uniformity for directed graphs.
However, the technique requires graphs to be acyclic, leading to similar problems with CFGs as traditional layered graph drawing algorithms.

IPSep-CoLa~\cite{dwyer_ipsep-cola_2006} goes in a similar direction, allowing for certain relative placement constraints between node pairs to be introduced.
Automatically deducing such constraints to enforce control flow-specific aesthetics criteria is challenging, making the technique less suitable for automatic CFG drawing.

Carmel et al.~\cite{carmel_combining_2004} introduced a technique where force minimization is combined with layered graph drawing, enabling drawing for both cyclic and acyclic graphs by not restricting nodes to strict ranks.
The approach is well suited for higher degree graphs, and does not aim to strictly enforce happens-before relationships, therefore being less suitable for control flow graphs.

\paragraph{Control Flow Graph Drawing}
Not many works have focused on improving graph drawing specifically for control flow graphs, but a few techniques have been put forth.

CFGExplorer~\cite{devkota_cfgexplorer_2018}, CFGConf~\cite{devkota_domain-centered_2022}, and CcNav~\cite{devkota_ccnav_2021} have addressed the issue of happens-before relationships being violated after with loops by forcing layered graph drawing algorithms to assign specific ranks to loop exits through invisible edges inserted during ranking.
While this improves on the happens-before criterion for loops, the remaining layout still relies on Dagre.
VEIL further improves on this by providing edge grouping and adding semantic meaning to edge lengths and bends.

\texttt{regVIS}~\cite{toprak_lightweight_2014} uses regular expression-based parsing of control flow graphs to extract common control flow constructs such as loops and visualize them in more intuitive manners, largely avoiding CFG layout issues.
The reliance on regular expressions means that the technique primarily handles regular / reducible control flow constructs.

Robillard and Simoneau~\cite{robillard_iconic_1993} have developed an alternative approach of visualizing control flow graphs using an iconic representation.
The technique avoids many layout related issues and produces slender, vertical control flow graph visualizations similar to VEIL, however with less strict adherence to other established graph aesthetics criteria.

In a direction similar to control flow graphs, Würthinger et al.~\cite{hendren_visualization_2008} have developed a system for visualizing program dependence graphs, which show program dependency relationships similar to control flow dependencies.
Balmas~\cite{balmas_displaying_2004} investigated a similar system for visualizing system dependence graphs.
Under the hood, both systems rely on Graphviz's \texttt{dot} layout, meaning cycles in control flow graphs lead to similar violations of happens-before relationships demonstrated in this paper.

\section{Conclusion}
We present VEIL, a novel graph drawing algorithm for control flow graphs that explicitly incorporates program semantics into the drawing process.
Unlike general-purpose graph drawing approaches, VEIL leverages dominator analysis and CFG-specific aesthetics criteria to produce layouts that better reflect execution order and structural program constructs.
In doing so, it codifies and extends established graph drawing aesthetics with two new domain-specific criteria: happens-before ordering (C10) and edge direction grouping (C11).

Through a systematic evaluation on real-world CFGs, we compare VEIL against state-of-the-art baselines in layered graph drawing across 11 quantitative metrics.
The results demonstrate that VEIL consistently improves adherence to CFG-relevant layout criteria while maintaining or exceeding performance on general-purpose graph drawing aesthetics.
In particular, VEIL provides clearer visual cues for loops, branching, and execution flow, thereby improving readability and enabling users to reason about program behavior more naturally, akin to reading source code.

Looking ahead, our work suggests several promising directions. Integration of VEIL into mainstream compiler toolchains and security analysis environments could enhance the usability of CFG visualizations in practice.
Further user studies would also complement our quantitative evaluation by assessing human comprehension directly.
Finally, the principles underlying VEIL may extend beyond CFGs to other hierarchical graph domains where execution order or semantic flow is central.

In summary, VEIL demonstrates that domain-aware layout algorithms can substantially advance the clarity and effectiveness of program visualizations, bridging the gap between formal graph structures and human program understanding.

\begin{acks}
This project has received funding from the European Research Council (ERC) under the European Union’s Horizon 2020 program (grant agreement PSAP, No. 101002047), and from the European High Performance Computing Joint Undertaking (EuroHPC-JU) under the DEEP-SEA program (grant agreement No. 955606).
Work by Lawrence Livermore National Laboratory was performed under the auspices of the U.S. Department of Energy under contract DE-AC52-07NA27344 (LLNL-CONF-2013055).
\end{acks}

\bibliographystyle{ACM-Reference-Format}
\bibliography{references, additional-refs}

\end{document}